\documentclass[12pt]{article}
\usepackage[toc,page]{appendix}
\usepackage{hyperref}
\usepackage{graphicx,pdflscape}
\usepackage{color}
\usepackage{bm}
\usepackage{alltt,dsfont}
\usepackage{appendix}
\usepackage{amsmath,amssymb}

\newcommand {\apgt} {\ {\raise-.5ex\hbox{$\buildrel\rangle\over\sim$}}\ }
\newcommand {\aplt} {\ {\raise-.5ex\hbox{$\buildrel\langle\over\sim$}}\ }

\newcommand{\tr}{{\text {Tr} }}

\newcommand{\si}{{\sigma}}
\newcommand{\sib}{{\bar{\sigma}}}

\newcommand{\beq}{\begin{eqnarray}}
\newcommand{\eeq}{\end{eqnarray}}
\newcommand{\barray}{\begin{eqnarray}}
\newcommand{\earray}{\end{eqnarray}}
\newcommand{\nn}{\nonumber}

\newcommand{\up}{\uparrow}
\newcommand{\dn}{\downarrow}

\renewcommand{\k}{ \vec{k}}

\newcommand{\pib}{\frac{\pi}{\beta}}
\newcommand{\tpib}{\frac{2 \pi}{\beta}}
\newcommand{\invbeta}{\frac{1}{\beta}}

\usepackage{hyperref}
\hypersetup{
    colorlinks,
    citecolor=red,
    filecolor=cyan,
    linkcolor=blue,
   urlcolor=magenta
 }

\usepackage{xparse}
\ExplSyntaxOn
\NewDocumentCommand{\mref}{m}{\quinn_mref:n {#1}}
\seq_new:N \l_quinn_mref_seq
\cs_new:Npn \quinn_mref:n #1
 {
  \seq_set_split:Nnn \l_quinn_mref_seq { , } { #1 }
  \seq_pop_right:NN \l_quinn_mref_seq \l_tmpa_tl
  ( 
  \seq_map_inline:Nn \l_quinn_mref_seq
    { \ref{##1},\nobreakspace } 
  \exp_args:NV \ref \l_tmpa_tl 
  ) 
 }
\ExplSyntaxOff

\newcommand{\disp}[1]{Eq.~\mref{#1}}

\newcommand{\figdisp}[1]{Fig.~\mref{#1}}

\newcommand{\de}{\delta_\varepsilon}
\newcommand{\half}{\frac{1}{2}}
\newcommand{\ubytwo}{\half U}

\usepackage{amsthm}
\theoremstyle{plain}
\newtheorem*{theorem*}{Theorem}
\usepackage{graphicx} 
\graphicspath{{./Figures/}} 

\begin{document}
\title{   Partition function zeros  of quantum many-body systems}
\author{ B Sriram Shastry$^{1}$\footnote{sriram@physics.ucsc.edu}  \\
\small \em $^{1}$Physics Department, University of California, Santa Cruz, CA, 95064 }
\date{ July 11, 2025}

\maketitle
\begin{abstract}
We present a {  new} method for calculating the Yang-Lee partition function zeros of a translationally invariant model of lattice fermions, exemplified by the Hubbard model.  The method rests on a  theorem
involving the  usual single electron self-energy $\Sigma_\si(\k, i \omega_n|\mu)$ with chemical potential $\mu$,   in the  imaginary time  Matsubara formulation. The theorem
  maps the Yang-Lee zeros  to  a set of  wavevector and spin labeled
virtual energies $\xi_{\k \si}$. These,  thermodynamically derived virtual energies, are  solutions of the set of   equations $\xi_{k \si}=\varepsilon_\si({\k})-\half U+ \Sigma_\si(\k, i \pib|\xi_{k \si}+\half U-i \pib) =0$.     Examples of the method in simplified situations  are provided. 

\end{abstract}


\section{Introduction}

C. N. Yang and T. D.  Lee launched an important and fruitful direction in condensed matter physics, by highlighting the significance of the zeros of the grand partition function of many-body systems\cite{Yang-Lee}.
  Lee and Yang\cite{Lee-Yang}, soon thereafter, showed that  for Ising ferromagnets, the zeros can be located exactly and  lie on the unit circle in the complex fugacity plane. They showed that the density of the zeros closest to the real line provides  fundamental insight into the critical singularities near a phase transition in 2 and higher dimensions. The  work of T. Asano\cite{Asano-Contractions} gave an elegant reformulation of the Lee-Yang results using the method of ``contractions'', and also enabled the surmounting of the added difficulties due to quantum mechanics in the ferromagnets\cite{Asano-Quantum}.
  The methodology used for these results was  streamlined by D. Ruelle and F. Dyson\cite{Ruelle}, leading to several important further results over the intervening years\cite{Lieb,Lebowitz}, reviewed in \cite{Review-YL}. The currently existing  methods unfortunately do not appear to be generalizable for antiferromagnets and  for broader classes of systems, including the ones considered here. 

Our interest is in the study of quantum many-body systems, such as the Hubbard model (see \disp{Hubbard-H}).
Locating the zeros and extracting meaningful information for these physically important systems  has proceeded at a sedate pace.
 Here the major difficulty is the absence of systematic analytical  methods. For numerical studies,  a combination of the  very high precision required for locating the zeros and the exponential growth of the Hilbert spaces with the system size, hamper direct enumeration techniques.  Quantum Monte Carlo (QMC) methods for studying the repulsive Hubbard model with a complex chemical potential have been used earlier \cite{Dagotto,Glasgow,Glasgow-2,Wilczek}. However  the full spectrum of zeros seems difficult  to obtain from QMC. {  Studies of related problems employing  tensor networks show promise in overcoming some of these limitations \cite{Kist,Liu}. } There have also been recent studies of partition function zeros for related quantum many-body models   in a variety of contexts, these include decoherence studies\cite{Wei}, quantum quenches\cite{quenches} and  in quantum computation \cite{Johri}. These  indicate a renewal of interest in this important subject.

The main objective of this paper is to introduce a new analytical method for locating the zeros of the grand partition function of the Hubbard model (see \disp{Hubbard-H})- in any dimension. This method  requires  a  knowledge of the  self-energy in the Matsubara imaginary time framework\cite{MATSUBARA}. While the self-energy is a highly non-trivial object to compute,
a perturbative knowledge of the self-energy is available and may be used to give approximate location of the zeros. 
The method generalizes to  nearby models in a straightforward way. The  a few simple  examples are provided below. 

\subsection{Heuristic perturbative approach to Yang-Lee zeros}
\begin{figure}[t]
\centering
\includegraphics[width=.45\columnwidth]{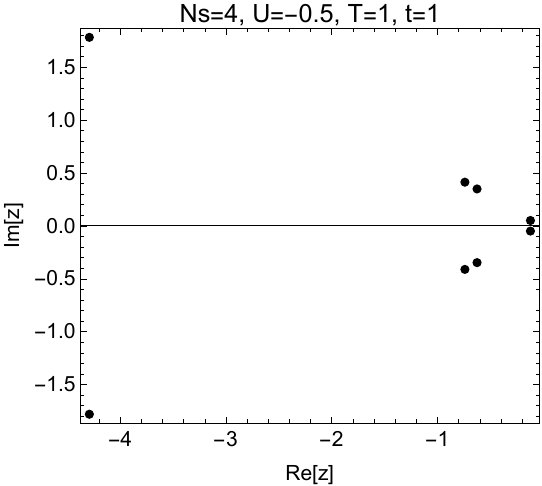}
\includegraphics[width=.45\columnwidth]{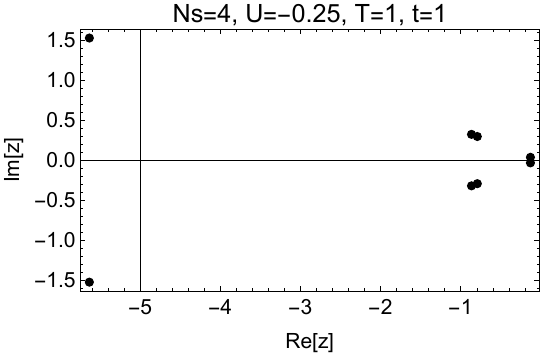}\\
\includegraphics[width=.45\columnwidth]{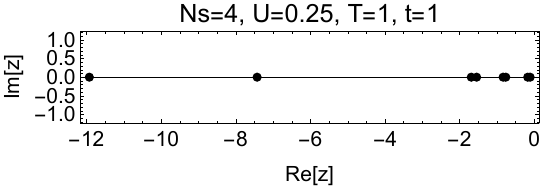}
\includegraphics[width=.45\columnwidth]{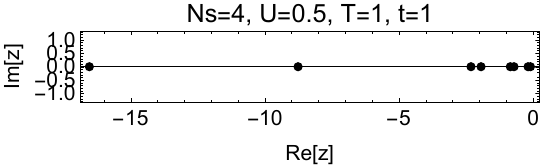}
 \caption{\footnotesize  Yang-Lee zeros of the partition function of the 4 site Hubbard model at T=1: {\bf TopLeft} with U=-0.5 {\bf TopRight} with U=-0.25 {\bf BottomLeft} with U=+0.25 and {\bf BottomRight} with U=0.5. The partition function was calculated using the symbolic calculation programme {\em DiracQ}\cite{DiracQ}.
 From these figures we see that the attractive case leads to complex fugacities lying on a parabolaic curve- while the repulsive case leads to real fugacities. This is as expected from  \disp{pert-1}. \label{Fig.1}}
 \end{figure}

Let us consider the Hubbard model \disp{Hubbard-H} as an illustrative example.
A  perturbative approach is to expand the (grand) partition function as a series in the coupling constant \cite{AGD}  $Z=\tr e^{- \beta ({\cal H}_0+V)}= Z_0 (1- \beta \langle T_\tau V \rangle_0) + {\cal O}(U^2) $, where $Z_0=\prod_\alpha (1+ z e^{-\beta \varepsilon_\alpha})^2$  is the non-interacting result with the fugacity $z=e^{\beta \mu}$, $T_\tau$ the imaginary-time ordering operator, $\beta=\frac{1}{k_B T}$  and the correction term due to interactions is found by taking the non-interacting thermal average of the interaction term, so that  
\beq
Z(z)= \prod_\alpha (1+ z e^{-\beta \varepsilon_\alpha})^2 \times \left[ 1-\frac{\beta U}{N_s} \left(\sum_{\alpha} \frac{z}{z+ e^{\beta \varepsilon_\alpha}}\right)^2 +{\cal O}(U^2)\right], \label{pert-0}
\eeq 
where the spin degeneracy leads to the square in the second term. For simplicity let us assume that all the energies $\varepsilon_\alpha$ are distinct. A simple minded estimate of the location of the zeros can be made as follows. Since the zeros of $Z_0$ are located at 
$z_\alpha \to - e^{\beta \varepsilon_\alpha}$, we  might expect that the zeros of $Z$ are continuously connected to these and hence try the expression $z_\alpha= - e^{\beta (\varepsilon_\alpha+\Delta_\alpha)}$ where the energy shift $\Delta_\alpha$ vanishes as $U\to0$. With a fixed $\alpha$ we substitute into  \disp{pert-0} and the vanishing condition imposed on $Z$, thus requiring
\beq
 \sum_{\gamma} \left(\frac{1}{1- e^{\beta (\varepsilon_\gamma-\varepsilon_\alpha-\Delta_\alpha)}}\right) = \pm \sqrt{\frac{N_s}{\beta U}}. \label{heur-1}
\eeq 
Solving  to low orders we get
\beq
\Delta_\alpha= \pm \sqrt{\frac{ U k_B T}{N_s}}+ \frac{U}{ 2 N_s} \sum_{\gamma \neq \alpha} \coth{\frac{\beta}{2}(\varepsilon_\gamma-\varepsilon_\alpha)} + \ldots  \label{pert-1}
\eeq
From this expression
we  infer that for $U>0$ the energy shifts are real, and the fugacities $z_\alpha$ are on the negative real axis. On the other hand for any $U<0$ the energy shifts are complex numbers occurring in complex conjugate pairs, and the fugacities $z_\alpha$ lie on a  curve in the negative half plane which is symmetric about the real axis. In \figdisp{Fig.1} we present the roots for the 4-site Hubbard model, where the partition function was calculated using the software package {\em DiracQ}\cite{DiracQ} employing symbolic methods. We see that the exact roots behave  in accordance with the above observation.
The change in the nature of roots- from real for repulsive to non-real for attractive interactions
is pronounced, and easily visible. This  provides a graphic demonstration of Dyson's comments\cite{Dyson} on  the nature of perturbation theory in many-body systems, and is discussed further below.

We therefore obtain some insight into the roots of the Hubbard model from the simple perturbation theory outlined above. However the theoretical argument is not completely satisfactory, since it mixes in factors of very different orders in $N_s$ in the expression \disp{pert-0}, or its higher order generalizations. It is highly desirable to have a formulation where the fugacity-which is of the ${\cal O}(1)$ in $N_s$- is found from equations where all terms are of the same, i.e. ${\cal O}(1)$ order. We provide  such a framework in this work. 

\subsection{Plan of the paper}
In Section(\ref{Section2}) we provide basic definitions of the Hamiltonian, the grand partition function and the fugacity. We define the virtual energies $\xi_{\k \si}$, and note some  properties of these obtained from general considerations. In Section(\ref{Section3}) we define the Matsubara finite temperature Greens functions, their un-normalized versions and the self-energy. In Section(\ref{Section4}) the main theorem of this work is proposed and established. The  equation   determining the virtual energy in terms of the self-energy is recorded in \disp{solver}. In Section(\ref{Section5}) we illustrate the application of the theorem to the problem of an exactly solvable single site Hubbard model. Here the role of lifting of  degeneracy of the single site energy is also studied in detail. The exact self-energy is expanded to ${\cal O}(U^2)$, and to this order the perturbative results for $\xi$ are shown to agree with the exact ones. This somewhat long section is mainly addressed to readers interested in details, and may be skipped by others. In Section(\ref{Section6}) we use the first order (Hartree) approximation to the self-energy and extract the virtual energies $\xi_{\k \si}$ to ${\cal O}(U)$ for both signs of the interaction for the 1-d Hubbard ring of length 6. This is familiar as the idealized Benzene molecule. These virtual zeros  are compared with exact numerical values obtained by computing the grand partition function exactly using symbolic methods. In Section(\ref{Section7}) we note a few conclusions and prospects for further work.

In Appendix(\ref{Section8}) we recall the  well known results of Luttinger and Ward\cite{Luttinger-Ward} for the grand partition function. Their result neatly expresses the result of an all-orders summation of  perturbation theory.  We observe that the key  function $\Xi(\mu)$ in \disp{our-main}- whose zeros are all the virtual energies-  is a multiplicative factor of the Luttinger-Ward  partition function.

\section{Basic Hamiltonian and the partition function \label{Section2}}
Let us consider an interacting fermi system described by the (grand-canonical) Hamiltonian 
\beq
{\cal H}& =& H - \mu \, {\cal N} = {\cal H}_0+ V \nn \\
{\cal H}_0&=&\sum_{\k_\alpha \si} \varepsilon_{k_\alpha,\si}\, C^\dagger_{\k_\alpha \si} C_{\k_\alpha \si}-\mu {\cal N}, \label{Hubbard-H-old}
\eeq
where the (conserved)  fermion number operator  ${\cal N}=\sum_{k,\si}C^\dagger_{k \si} C_{k \si}$,  $\mu$  the chemical potential,  and the  Hubbard Hamiltonian 
\beq { H} = \sum_{\k_\alpha \si} \varepsilon_{k_\alpha,\si}\, C^\dagger_{\k_\alpha \si} C_{\k_\alpha \si} + U \sum_i n_{i \uparrow} n_{i \downarrow}. \label{Hubbard-H} \eeq 
Here $\k_\alpha$ are the $L^d$ box-quantized wave vectors.
The kinetic energy term  is the Fourier transform of the hopping Hamiltonian $T=- \sum_{\vec{r},\si,\vec{\eta}}(t_{\vec{\eta},\si} C^\dagger_{\vec{r} \si}C_{\vec{r}+\vec{\eta} \si }+ h.c.)$, and $\vec{\eta}$ is the nearest neighbour vector on  a d-dimensional hypercubic lattice. In the usual case the hopping matrix element $t_{\vec{\eta},\si}$ is independent of the spin $\si$. The generalized $\si$ dependent  case is useful for studying the effect of lifting the spin-degeneracy on the roots as discussed  below.  More general models can be obtained by replacing the interaction term with an appropriate $V$ and the energy dispersion chosen suitably. 
We have  assumed that the system has translation invariance- and study  a finite lattice (with lattice constant unity) with $N_s=L^d$ sites.

 For the  finite system under consideration the total number of (spin half) particles varies in the range  $0\leq N \leq 2 N_s$.    With the Boltzman constant $k_B=1$ and $\beta = \frac{1}{T}$
 we trace over all particle numbers and  define the  grand partition function:
\beq
Z(\mu) = \tr \,  e^{- \beta (H - \mu {\cal N})} = \sum_{n, N } e^{- \beta E_n+ \beta \mu N_n}= \sum_{N=0}^{ 2 N_s} z^{  N} Z_N(\beta), \label{partition-function}
\eeq
where  $E_n$ and $N_n$ are the eigenvalues of $H$  and particle number for the state $n$ (detailed definitions are given after \disp{Gbar}), $Z_N$ is the canonical partition function for $N$ particles and  $z $ is the fugacity $
 z(\mu)= e^{\beta \mu}. $ 
 $Z(\mu)$ is thus a polynomial in the fugacity $z $ of degree $ 2N_s$, and  by the fundamental theorem of algebra, it has exactly $2 N_s$ zeros in the complex $z$    plane- and hence expressible as
 $Z(\mu)=c_0 \prod_{\alpha=1}^{2 N_s} (z-z_{\alpha})$, where $z_\alpha$ are
  the Yang-Lee zeros of the given model.  We will find it more convenient to work in the 
 complex $\mu$ plane rather than with the fugacity $z$.  We note the periodicity $z(\mu)=z(\mu+ i \tpib)$ and therefore  the complex $\mu$ plane can be decomposed into infinite  equivalent  strips of width $i \tpib$. We  restrict our attention to a  conveniently chosen  strip
 \beq
 {\cal S} = \left\{ -  \tpib  < \Im m \,  \mu \leq  0, \;\; \Re e \, \mu  \in  [-\infty,\infty] \right\}. \label{mudomain}
 \eeq
 We remark further on the rationale underlying this convention  below \disp{conventionE}.


\subsection{Definition of the virtual energies $\xi_{\k \si}$}
We will write the partition function in the form 
\beq
Z(\mu)= \prod_{\vec{k} \si} (1+\frac{e^{\beta \mu}}{e^{\beta (\xi_{\vec{k} \si}+\ubytwo)}} ). \label{Eq-21}
\eeq
 With this convention- explained below- the roots of $Z$ occur at 
\beq
\mu&\to& - i \frac{\pi}{\beta}+\ubytwo+ \xi_{\vec{k} \si}, \mbox{ or equivalently }\nn \\
z_\alpha&\to&-e^{\beta (\xi_{\vec{k} \si}+\ubytwo)} \label{EYL}
\eeq
where \disp{EYL}  anticipates  a mapping $\alpha \leftrightarrow \k\si$, between the unlabeled Yang-Lee zeros $z_\alpha$ and the labeled virtual energies $\xi$.
We remark that while \disp{Eq-21} has the same form as that of an ideal Fermi gas with energies  $\xi_{\k \si}+\ubytwo$, the latter are non-trivial complex functions of $T, U$. For an interacting system $\xi_{\k \si}$ are not viewed as eigenvalues of any Hermitean Hamiltonian, and are allowed to be non-real. For this reason these may be regarded as {\em thermodynamically derived virtual energies}.

The  imaginary part of $\xi_{\k \si}$ is constrained by
\beq
-\pib < \Im m \, \xi_{\k \si} \leq  \pib. \label{conventionE}
\eeq
The convention used in defining the strip ${\cal S}$  in \disp{mudomain}, and the range of the imaginary part of $\xi$ in \disp{conventionE},  are  such that complex conjugate  roots in the fugacity variable $z=e^{\beta \mu}$  lead to complex conjugate  energies $\xi$. The partition function being a polynomial in $z$ with real coefficients, the roots in $z$ are either real or come in complex conjugate pairs. It follows that the virtual energies are either real or they occur in complex conjugate pairs, i.e. if $\xi$ is a virtual energy, then so is $\xi^*$.

The virtual energies
$\xi$'s are dependent on the temperature $T$, and also the interaction strength $U$. We also remark that if the 
virtual energies $\xi$ are  real  and their density of states is non-singular in a continuous  interval  of energies, then the system would behave like an almost ideal Fermi liquid for densities where $\mu$ lives in the same range. Such a situation seems to be realized in the repulsive Hubbard model at low $T$ studied here. It is tempting to view Landau's Fermi liquids \cite{Landau, Nozieres,AGD} as behaving in precisely such a fashion, with the energies $\xi_{\vec{k} \si}$ being closely related to Landau's quasiparticle energies \cite{Landau}.

We next record  some general properties of the virtual energies for the Hubbard model.
\begin{itemize}
\item On bipartite lattices there is an invariance under   a simultaneous particle hole transformation for both directions of the spin, it is expressible in real space as $C_{i \si} \leftrightarrow e^{i \pi \phi_i}  C^\dagger_{i \si}$  with a suitable phase factor $\phi_i$.
Collecting the extra coefficients generated by the transformation, we obtain  an identity
\beq
{ Z}(\mu) = e^{2 \beta \mu  N_s} e^{- \beta U N_s} {Z}(U-\mu) . \label{p-h-symmetry}
\eeq
This implies that the partition function zeros arise in particle-hole symmetric pairs $\{ \mu, U-\mu\}$, or equivalently the   virtual energies come in pairs, i.e.
\beq
 \xi_{k \si}= - \xi_{k' \si} , \label{particle-hole}
\eeq
for suitable momentum partners $k$ and $k'$.
\item In the fully occupied subspace (i.e. particle number $N= 2 N_s$ sector), there is only a single state. Here the kinetic energy is inoperative due to the ``jammed'' nature of the state, and hence it is simple to compute the corresponding canonical partition function $Z_{2 N_s}= e^{- \beta U N_s}$. Comparing with the coefficient of $e^{\beta N_s\mu}$ in \disp{Eq-21}, we find the first sum-rule:
\beq
 \sum_{\k \si} \xi_{\k \si}= 0. \label{sumrule-1}
\eeq
 The imaginary parts of $\xi$, if present, cancel out by the complex conjugate property of the roots. It is readily seen  that \disp{sumrule-1} is satisfied when the $\xi$'s satisfy the condition of particle hole symmetric pairs \disp{particle-hole} by summing over all $\{ \k \si \}$.

\item A second sum-rule can be readily found by comparing the results of the single particle sector $N=1$, i.e the coefficient of $e^{\beta \mu}$ in \disp{Eq-21}. Its  exact value is found using  the observation that the  two-particle interaction $U$ is inoperative in this sector. We find the second sum-rule:
\beq
e^{- \beta \half U} \sum_{\k \si} e^{- \beta  \xi_{\k \si}}=
\sum_{\k \si} e^{- \beta  \varepsilon_{\k \si}}\label{sumrule-2}.
\eeq
{  We now define $U_{eff}$ as   a function of the parameters $U,\beta$ and also the computed $ \{\xi\}$. It is given by
\beq
U_{eff}(U,\beta,\{\xi\})= \frac{2}{\beta} \log \left(\frac{\sum_{k \si} e^{- \beta \xi_{k \si}}}{\sum_{k \si} e^{- \beta \varepsilon_{k \si}}}\right). \label{Ueff}
\eeq
 The  sum-rule \disp{sumrule-2} may be cast in the form
\beq U_{eff}(U,\beta,\{\xi\})= U \label{Ueff-2}. \eeq
Clearly if the $\xi$'s are exactly found, then the sum-rule is satisfied exactly. If $\xi$'s are only found  approximately, then the quality of the   approximation   is reflected in the difference between $U_{eff}$  and $U$. This idea is  illustrated below in \figdisp{Figure3,Figure4}. }

\item Further sum-rules can be found by comparing other powers of $e^{\beta \mu}$, but they  require exact knowledge of  the canonical partition functions $Z_2,Z_3 \ldots $, and  are therefore less tractable.

\end{itemize}

\section{The Greens function and Self-energy in Matsubara frequencies \label{Section3}}
 
It is useful to consider  the {\em unnormalized} thermal Greens function, signified by an overbar 
\beq \bar{G}_\si(\k,\tau|\mu) = - \left(  \tr \, e^{- \beta {\cal H}} \left(T_\tau C_{\k \si}(\tau) C^\dagger_{\k \si}(0)\right) \right), \label{G}
\eeq 
where  $-\beta < \tau \leq \beta$, and  $T_\tau$ denotes the (Fermionic) imaginary-time ordering operator.  For clarity we display the dependence on the chemical potential for $G,\Sigma, \ldots$, as in \disp{G}. For any operator ${\cal Q}$   the standard imaginary time Heisenberg picture time dependence   is defined through ${\cal Q}(\tau) = e^{\tau {\cal H}} {\cal Q} e^{-\tau {\cal H}}$.
 This Greens function is related to the conventional normalized Greens function $G$ \cite{MATSUBARA,AGD} through
\beq
\bar{G}_\si(\k,\tau|\mu)= Z(\mu) G_\si(\k,\tau|\mu) \label{barG}.
\eeq
The usual antiperiodicity extends to the unnormalized $\bar{G}_\si(\tau+ \beta) = -\bar{G}_\si(\tau)$, and  used to obtain the Matsubara Fourier representation \cite{MATSUBARA} $\bar{G}_\si(\k, i \omega_r|\mu) = \frac{1}{2 \beta} \int_{- \beta}^\beta e^{i \omega_r \tau} \bar{G}_\si(\k, \tau|\mu) d \tau$, where the allowed Matsubara frequencies are  $\omega_r = (2 r+1) \pib$ with integer $r$. 
We next record  the  equation of motion for $\bar{G}$ following from the Heisenberg dynamics $\partial_\tau C_{\k,\si}=[{\cal H},C_{k\si}(\tau)]$
\beq
(- \partial_\tau + \mu- \varepsilon_\si(k) ) \bar{G}_\si(k,\tau|\mu) = \delta(\tau) Z(\mu) + \bar{\cal A}_\si(\k,\tau|\mu) \label{derivative-G-2}
\eeq
where
\beq
\bar{\cal A}(\k,\tau|\mu) = -   \tr \, e^{- \beta {\cal H}} \left(T_\tau a_{\k \si}(\tau) C^\dagger_{\k \si}(0)\right)  \label{derivative-G}
\eeq
with 
\beq a_{\k \si}\equiv [ C_{\k \si},V] . \eeq 
Note that we can also write the un-normalized $\bar{\cal A}=Z(\mu) {\cal A}$, in analogy with the corresponding G's.
Taking the Matsubara-Fourier series of both sides we get 
\beq
\left\{ i \omega_r + \mu- \varepsilon_\si(k)\right\}\bar{G}_\si(\k,i \omega_r|\mu)=Z(\mu)+ \bar{\cal A}_\si(\k, i \omega_r|\mu). \label{dyson-1}
\eeq
   Defining  the Dyson self-energy $\Sigma$ as 
\beq
 \Sigma_\si(\k, i \omega_r|\mu)= \frac{\bar{\cal A}_\si(\k, i \omega_r|\mu)}{ \bar{G}_\si(\k, i \omega_r|\mu)} \label{dyson-2}
\eeq 
this leads to the un-normalized version of  Dyson's equation
\beq
\left\{ i \omega_r + \mu- \varepsilon_\si(k)- \Sigma_\si(\k, i\omega_r|\mu)\right\}\, \bar{G}_\si(\k,i \omega_r|\mu)=Z(\mu). \label{main-eq}
\eeq
It should be noted that in  $\Sigma_\si(\k, i\omega_r|\mu)$, every occurrence of $i \omega_r$ is in  the combination $i \omega_r+\mu$. 
 In terms with $r \neq 0$ we note that $\omega_r-\omega_0=r \tpib$,  this shift can be absorbed into $\mu$ and using the periodicity $Z(\mu)=Z(\mu + i  r \tpib)$. We also note that \disp{eq14,dyson-1,dyson-2} implies  the periodicity 
 in the complex $\mu$ plane for integer `$s$'
 \beq
\bar{G}_\si(\k,i \omega_r|\mu+ i s \tpib)=\bar{G}_\si(\k,i \omega_{r+s}|\mu) , \label{G-periodic-in-mu}
 \eeq
 and similar relations for $\Sigma_\si(\k, i \omega_r)$.
  Therefore studying $\bar{G}_\si(\k, i \omega_r|\mu) $  with  $r\neq0$ does not lead to anything new and  in the following we  confine ourselves to $r=0$. Displaying the $\mu$ dependence of $G$ and $\Sigma$ explicitly  and defining $\Psi$ through
 \beq
 \Psi_{\k \, \si}(\mu)&\equiv& G^{-1}_{\si}(\k, i \pib|\mu)\nn \\
  &=& \mu+ i \pib- \varepsilon_\si({\k})- \Sigma_\si(\k, i \pib|\mu) ,
  \label{Psi}
 \eeq
we write \disp{main-eq} as 
  \beq
 \Psi_{\k \, \si}(\mu) \times \bar{G}_\si(\k, i \pib|\mu)= Z(\mu) \label{YLDyson}.
   \label{Condition}
 \eeq
 The total number of such equations is $2 N_s$, including the 2 spin projections  and the $N_s$ values of $\vec{k}$.

\section{Theorem for Locating the Yang-Lee  zeroes of $Z(\mu)$.  \label{Section4}}
We now  discuss the connection between Yang-Lee zeros of $Z(\mu)$ and the vanishing of $\Psi_{\k \si}(\mu)$ (see \disp{Psi}). Recalling \disp{mudomain} we will establish below  
 \begin{theorem*}
 For  a fixed  $(\k \, \si)$, if a complex number  $\mu^*\in {\cal S}$ is a zero of $\Psi_{\k \si}(\mu)$  (i.e. $\Psi_{\k \si}(\mu^*)=0$), then $\mu^*$ is also a zero of the partition function (i.e. $Z(\mu^*)=0)$. 
  \label{HAT}
 \end{theorem*}

{\bf Remarks:} We  record a few relevant comments below.
\begin{itemize}

\item   In principle the self-energy $\Sigma_\si(\k,i \omega_r|\mu)$  can be calculated to arbitrary  orders in the interaction, using the standard rules of the   Feynman-Dyson-Matsubara perturbation theory in the imaginary time\cite{MATSUBARA,AGD,Luttinger-Ward}.
 Given an approximation  for $\Sigma$ to a certain order in the coupling, the
{ theorem } can be used to find the  corresponding momentum-labeled virtual energy $\xi_{\k \si}$, and from their total set, the (unlabeled) Yang-Lee zeros of the partition function.

\item If  the non-interacting virtual energies (i.e. the band energies $\varepsilon_{\k \si}$) are completely non-degenerate, then using continuity in the coupling,   {\em all zeros} of $Z$ can be located perturbatively from the { theorem} by varying $\k,\si$.

\item    \disp{Condition} also admits  the  possibility that $\mu^*$, a root of $Z$, satisfies  $\bar{G}_\si(\k, i \pib|\mu^*)=0$ for a fixed $\k \si$, while $\Psi_{\k \si}(\mu^*)\neq0$. This type of root is much less amenable to a systematic perturbative treatment since $\bar{G}$, unlike $\Psi_{\k \si}$, is generally  a function of almost the entire set of $\k \si$.   However we may bypass this class of roots- by studying the union over the $\k,\si$ of the  class of roots    for each individual  $\k,\si$'s.

\item  With the non-degeneracy assumption noted above,  our result is that  all zeros of $Z$ coincide with the zeros of a function $\Xi(\mu)$ defined  by
 \beq
\Xi(\mu)= \prod_{\k \si}  \Psi_{\k \si}(\mu) \label{our-main}.
\eeq
\item When the single electron levels $\varepsilon_{\k \si}$ have degeneracies due to spin invariance or parity, this case can be viewed as a limiting case of the non-degenerate one. In the discussion following \disp{eq64} we remark on how the degenerate case, treated to lowest order in $U$, allows for two perturbed solutions arising from a single (unperturbed) degenerate one, so that one can assign them one virtual energy to each spin projection. 
\item Using \disp{EYL} and \disp{Psi} we summarize the final equation determining the virtual energy $\xi_{\k \si}$ as
\beq
\xi_{\k \si}= \varepsilon_{\k \si}- \ubytwo + \Sigma_{\si}(\k, i \pib| \xi_{k \si} + \ubytwo- i \pib) \label{solver}.
\eeq
\end{itemize}

{\bf Proof}: To  prove the {\bf theorem}, a glance at \disp{Condition} shows that it is sufficient to show  that $\bar{G}$ is a holomorphic function  of $\mu$  in the strip ${\cal S}$ (\disp{mudomain}) (i.e. without any  poles or other singular points).
We can establish this property using the standard eigenbasis representation
\beq
\bar{G}_\si(\k,i \omega_r|\mu)= \sum_{n,m} \frac{e^{- \beta \epsilon_n}+e^{- \beta \epsilon_m} }{\epsilon_n-\epsilon_m + i \pib (2 r+1)} |\langle n|C_\si(\k)|m\rangle|^2 \label{Gbar}
\eeq
where  $\left\{ | n  \rangle \right\} $ denotes a complete set of (Fock) states ${\cal H}|n\rangle=\epsilon_n |n\rangle,   H |n\rangle= E_n |n\rangle, {\cal N} |n\rangle= N_n |n\rangle$, so that $\epsilon_n=E_n- \mu N_n $.  This spectral representation follows from \disp{G} by inserting complete sets of eigenstates followed by a Fourier transform from $\tau$ to the Matsubara frequencies $ \omega_r$.
Recall that the destruction operator $C_\si$ acting to the right either destroys the state or decreases the number of particles by 1, i.e.  $N_n-N_m=-1$. Making the $\mu$ dependence explicit we write
 \beq
 \frac{e^{- \beta \epsilon_n}+e^{- \beta \epsilon_m} }{\epsilon_n-\epsilon_m + i \pib(2 r+1)}=e^{ \beta (\mu N_n- E_n)}  \frac{1+e^{ \beta (E_n-E_m+\mu)} }{E_n- E_m +\mu+ i \pib (2 r+1)}, \label{eq14}
 \eeq
 and hence
 \beq
 \bar{G}_\si(\k, i \pib|\mu)= \sum_{n,m}e^{ \beta (\mu N_n- E_n)}  \frac{1+e^{ \beta (E_n-E_m+\mu)} }{E_n- E_m +\mu+ i \pib } |\langle n|C_\si(\k)|m\rangle|^2.
 \eeq
 The reality of the eigenvalues $E_n$ shows that $\bar{G}$  has no poles in the strip \disp{mudomain} due to the cancellation of a possible pole at $E_n- E_m +\mu+ i \pib =0$ by the vanishing of the numerator $1+e^{ \beta (E_n-E_m+\mu)}$  at the same location. Noting also that $0\leq |\langle n|C|m\rangle|^2 \leq 1$, $\bar{G}$ is therefore  a holomorphic  function of $\mu$ in the strip ${\cal S}$. From \disp{Condition} we obtain the { theorem}.

 \subsection{Another proof of \disp{main-eq,Condition}}

 We note a spectral representation for $\bar{A}$ (of the same form as \disp{Gbar})
 \beq
\bar{A}_\si(\k,i \omega_r)= \sum_{n,m} \frac{e^{- \beta \epsilon_n}+e^{- \beta \epsilon_m} }{\epsilon_n-\epsilon_m + i \pib (2 r+1)} \langle n|a_\si(\k)|m\rangle \langle m|C^\dagger_\si(\k)|n\rangle \label{Abar}.
\eeq
Using this relation and \disp{Gbar} we rearrange \disp{dyson-1}    with $r=0$ as
\beq
Z(\mu)&=& \left\{ i \pib +\mu - \varepsilon_{\si}(\k)\right\}\sum_{n m} \frac{e^{- \beta \epsilon_n}+e^{- \beta \epsilon_m} }{\epsilon_n-\epsilon_m + i \pib} |\langle n|C_\si(\k)|m\rangle|^2  \nn \\
&& - \sum_{n,m} \frac{e^{- \beta \epsilon_n}+e^{- \beta \epsilon_m} }{\epsilon_n-\epsilon_m + i \pib} \langle n |a_\si(\k)|m\rangle \langle m|C^\dagger_\si(\k)|n\rangle \label{Eq-18}
\eeq
which can be simplified using
\beq
\langle n|a_\si(\k)|m\rangle&=&\langle n[C_\si(\k),{\cal H]}|m\rangle-(\varepsilon_\si(\k)-\mu)\langle n|C_\si(\k)|m\rangle,\nn \\
&=&(\epsilon_m-\epsilon_n +\mu- \varepsilon_\si(\k)) \langle n|C_\si(\k)|m\rangle. \label{Eq-19}
\eeq
so that combining the two terms \disp{Eq-18,Eq-19},  and simplifying gives
\beq
Z(\mu)&=& \sum_{nm} \sum_{n m} (e^{- \beta \epsilon_n}+e^{- \beta \epsilon_m} ) |\langle n|C_\si(\k)|m\rangle|^2 \nn\\
&=& \mbox{Tr} e^{-\beta {\cal H}} \{ C_\si(\k),C^\dagger_\si(\k) \}\nn \\
&=&   \mbox{Tr} e^{-\beta {\cal H}},
\eeq
i.e. we recover the exact definition of $Z$ \disp{partition-function}. Summarizing, we see that  \disp{main-eq,Condition}  consistently represents  the vanishing of the partition function $Z(\mu)$ \disp{partition-function}.

\section{ Some illustrative examples \label{Section5}}
\subsection{The Non-interacting Fermi system}
We note that the non-interacting Fermi model with a partition function
\beq
Z_{non}(\mu)= \prod_{\vec{k} \si} \left(1 + \frac{e^{\beta \mu}}{e^{\beta \varepsilon_\si(\k)}} \right), \label{free-part}
\eeq
provides a simple illustration of our ideas, since a vanishing condition of the non-interacting Fermion Green's function
\beq
G^{-1}_{(0) \si}(\k, i \pib|\mu)= {i \pib + \mu - \varepsilon_\si(\k)},
\eeq
gives a zero of $Z_{non}(\mu)$. The added imaginary part $i \pib$  to   $\mu$ converts the Fermi factor into a Bose factor. It is therefore amusing to note that    root finding is performed on the ``Bosonized'' partition function.

\subsection{Exact Green's function in the Atomic Limit}
We study a less trivial  example of a single site Hubbard model in the so called atomic limit, where electrons are localized on an atom. Here the partition function, its roots and the exact Greens function are all calculable explicitly, and show that the { theorem} applies well to the exact solution for $G$ and also enables us to check the perturbative solution- which is carried out to ${\cal O}(U^2)$.

We consider the atomic limit of the Hubbard model with spin dependent energies $\varepsilon_{i \si}$ 
\beq 
H=\sum_{i \si} \varepsilon_{i \si} C^\dagger_{i \si} C_{i \si} + U \sum_i C^\dagger_{i \up} C_{i \up} C^\dagger_{i \dn} C_{i \dn}, 
\eeq
which is uncoupled in the sites  ``i"  and therefore easily solvable.
Let us first consider the partition function which is easily evaluated in terms of the fugacity $z$ and other parameters as
\beq
Z(\mu)&=&\prod_{i=1,N_s} {\cal Z}_i(\mu), \nn \\
{\cal Z}_i(\mu)&=& 1+ z (e^{-\beta \varepsilon_{i \up}}+e^{-\beta \varepsilon_{i \dn}})+z^2 e^{- \beta ( U + \varepsilon_{i \up} +\varepsilon_{i \dn})}.
\eeq
We simplify the notation using $\varepsilon_{i \si}=\varepsilon_{i 0}+\si \delta_\varepsilon$, so that
\beq
{\cal Z}_i(\mu)&=& 1+ 2 z e^{- \beta \varepsilon_{i 0}} \cosh \beta \de+z^2 e^{- \beta ( U + 2 \varepsilon_{i 0})},
\eeq
so that the two roots of ${\cal Z}_i(\mu)$ in the fugacity variable $z=e^{\beta \mu}$, which are  denoted as  $z_{i \si }$, can be calculated  as
\beq
z_{i \si }= -1\times e^{ \beta (U + \varepsilon_{i 0})} \left(\cosh \beta \de + \si  \{\sinh^2(\beta \de) +1 -e^{-\beta U }\}^\half \right).
\eeq 
At $U=0$ the roots  are $z_{i \si}=-e^{\beta \varepsilon_{i \si}}$ the free particle roots, while for non-zero $U$ the roots are either  on the negative real $z$ line ($U$ positive) or off the $z$ real line ($U$ negative). We can also express this result in terms of virtual energies $\xi_{i \si }$ defined in \disp{EYL} as
\beq
\xi_{i \si }= \ubytwo + \varepsilon_{i 0}+ \frac{1}{\beta} \log  \left(\cosh \beta \de + \si  \{\sinh^2(\beta \de) +1 -e^{-\beta U }\}^\half \right) \label{EYL2}
\eeq

We now drop the site index $i$, and compute the average number of particles using $n_\si=-\frac{1}{\beta} \frac{\partial}{\partial \varepsilon_\si } \log {\cal Z}$
\beq
n_{\si}(\mu)=\frac{1}{{\cal Z}(\mu)} \left(e^{\beta (\mu- \varepsilon_\si)}+ e^{\beta ( 2 \mu -U- 2 \varepsilon_0)}\right) \label{nsigma}
\eeq

Now we write down the exact atomic limit  Greens function- known from Hubbard's work\cite{Hubbard}
\beq
G_{\si}(i \omega_n|\mu)= \frac{1-n_{\sib}(\mu)}{i \omega_n+\mu- \varepsilon_\si}+\frac{n_{\sib}(\mu)}{i \omega_n+\mu- U-\varepsilon_\si},\label{GAtomic}
\eeq 
which can  be rewritten in the form of a single term with a self-energy
\beq
G_{\si}(i \omega_n|\mu)&=& \frac{1}{i \omega_n + \mu - \varepsilon_\si- \Sigma_\si(i \omega_n|\mu)}, \nn \\
\Sigma_\si(i \omega_n|\mu)&=& U n_{\sib} + \frac{U^2 n_{\sib}(1-n_\sib)}{i \omega_n+\mu -\varepsilon_\si- U(1-n_{\sib} )}
\eeq
We   alternately calculate $n_\si$ from \disp{GAtomic} using the familiar formula $n_\si= \frac{1}{\beta} \sum_n e^{i \omega_n 0^+} G_\si (i \omega_n)$.
After some algebra one can verify that the result is the same as in \disp{nsigma}.

Note also that the un-normalized Greens function \disp{barG}  is given by
\beq
\bar{G}_\si(i \omega_n|\mu)=\frac{1+e^{\beta(\mu-\varepsilon_\si)}}{i \omega_n+\mu- \varepsilon_\si}+\frac{e^{\beta(\mu-\varepsilon_\sib)}\left( 1+e^{\beta(\mu-\varepsilon_\si-U)}\right)}{i \omega_n+\mu-\varepsilon_\si- U}\label{barGAtomic}.
\eeq
It is easily seen that $\bar{G}_\si(i \omega_0)$ is a holomorphic function of $\mu$ in the strip ${\cal S}$ \disp{mudomain}, due to the vanishing of the numerator at every  location of the vanishing denominators.

\subsubsection{ Roots for small U: Non degenerate case $\de>0$}

We find by expanding \disp{EYL2}
\beq
\xi_\up= \epsilon_0+\de+ \ubytwo {\coth \beta \de}- \beta U^2 \frac{1}{8} \frac{\cosh \beta \de}{\sinh^3 \beta \de}+ {\cal O}(U^3) \label{EYLdirect}
\eeq

We can find  $\xi_\dn$ from the above by  setting $\de\to - \de$.

\subsubsection{Roots for small U: Degenerate case $\de=0$}
Expanding the $\de=0$ limit of \disp{EYL2} we get
\beq
\xi_+&=& \varepsilon_0- \sqrt{\frac{U}{\beta}}+\frac{\sqrt{\beta}}{12} U^{\frac{3}{2}}+ {\cal O}(U^{\frac{5}{2}}) \nn \\
\xi_-&=& \varepsilon_0- \sqrt{\frac{U}{\beta}}-\frac{\sqrt{\beta}}{12} U^{\frac{3}{2}}+ {\cal O}(U^{\frac{5}{2}})  \label{EYLdirect-deg}
\eeq 

\subsubsection{Perturbative Roots from $G$-The non-degenerate case}
We next show how the small U roots given in \disp{EYLdirect,EYLdirect-deg} can be obtained using the perturbative expansion of the Dyson self-energy, as remarked below the { theorem}, i.e.  the condition  $\Psi_{\k \si}(\mu)=0$.
Recall that
\beq
G^{-1}_\si(i \pib|-i \pib+ x)=x- \varepsilon_\si-\Sigma_\si(x) \label{gx}
\eeq
with
\beq
\Sigma_\si(x)=  U n_{\sib}+ \frac{U^2 n_{\sib}(1-n_\sib)}{x - \varepsilon_\si- U(1-n_{\sib} )}.
\eeq
For the purpose of this calculation we use the convenient variable $x$ which is related to the virtual energy through $\xi=x-\ubytwo$.
We next expand the exact  occupation $n_\si$ appearing above, using   \disp{nsigma}, in a series in $U$.
In terms of the Fermi function 
\beq
f_{z,\si}&=&\frac{z}{z+e^{\beta \varepsilon_\si}} 
\eeq
we find
\beq
n_\si&=& f_\si - \beta U f_\si f_\sib (1-f_\sib) +{\cal O}(U^2)\label{nexpand} 
\eeq
where $\sib=-\si$.
We setup the root equation to ${\cal O}(U^2)$
for $G_\si^{-1}=0$. Recall we are temporarily using $ x=i \frac{\pi}{\beta}+\mu$, so that
$z\to -e^{\beta x}$, and in terms of $x$
\beq
f_{z,\si}\vert_{z\to -e^{ \beta x}}&\equiv&f_\si(x)=\frac{e^{\beta x}}{e^{\beta x}-e^{\beta \varepsilon_\si}} \nn \\
f'(x)&=&\beta f_\si(x) (1-f_\si(x))
\eeq

Expanding $n_\si$ using \disp{nexpand}, and  truncating to ${\cal O}(U^2)$ or from diagrammatics directly, we get
\beq
\Sigma_\si(x)=  U f_{\sib}- \beta U^2 f_\sib(x)(1-f_\sib(x)) f_\si(x) + \frac{U^2 f_{\sib}(1-f_\sib)}{x - \varepsilon_\si} +{\cal O}(U^3)
\eeq
In \disp{gx} we substitute the above, and so the condition for a root is 
\beq
x-\varepsilon_\si -U f_\sib(x) +\beta U^2 f_\si(x) f_\sib(x) (1-f_\sib(x)) -U^2 \frac{f_\sib(x)(1-f_\sib(x))}{(x-\varepsilon_\si)}=0 \label{root-condition}
\eeq
and for therefore for  $\si=\up$
\beq
x&=&\varepsilon_\up +U f_\dn(x) -\beta U^2 f_\up(x) f_\dn(x) (1-f_\dn(x) ) +U^2 \frac{f_\dn(x)(1-f_\dn(x))}{x-\varepsilon_\up}=0\nn \\
&=&\varepsilon_\up + U f_\dn(x)- U^2 \beta f_\dn(x)(1-f_\dn(x)) \left[ f_\up(x)-\frac{k_B T}{x-\varepsilon_\up} \right]
\eeq

We can solve iteratively
\beq
x= x_0+ U x_1 + U^2 x_2+\ldots \label{xall}
\eeq
so that 
\beq
x_0&=&\varepsilon_\up+\eta= \varepsilon_0+\de +\eta \label{x0} \\
x_1&=& f_\dn(x_0) =  \frac{e^{\de}}{2 \sinh \beta \de} \label{x1}
\eeq
where we set $\eta=0$ at the end.
To get the full second order result, we need to expand
\beq
f_\dn(x_0+U x_1)&=&f_\dn(x_0)+ U x_1 f'_\dn(x_0) +.. \nn \\
&=&f_\dn(x_0)+\beta U f^2_\dn(x_0) (1-f_\dn(x_0))+.. 
\eeq
Hence
\beq
x_2=\beta  f^2_\dn(x_0) (1-f_\dn(x_0))- U^2 \beta f_\dn(x_0)(1-f_\dn(x_0)) \left[ f_\up(x_0+\eta)-\frac{k_B T}{x_0-\varepsilon_\up+\eta} \right]
\eeq
The term in square brackets  on the right equals $\half$  by taking the limit $\eta\to0$. On simplification this leads to
\beq
x_2= - \frac{\beta}{8} \frac{\cosh \beta \de}{ \sinh^3(\beta \de)} \label{x2}
\eeq 
We therefore verify that \disp{x0,x1,x2} combine to give the first three terms obtained directly \disp{EYLdirect}. 

\subsubsection{Degenerate case}
For the degenerate case $\varepsilon_\si=\varepsilon_0$ (i.e. $\delta_\varepsilon=0$), we show how the leading behaviour in \disp{EYLdirect-deg} is obtained from  the condition  $\Psi_{\k \si}(\mu)=0$ from the { theorem}. We write the leading, i.e. ${\cal O}(U)$ terms of \disp{{root-condition}}
\beq
x-\varepsilon_0= U\frac{1}{1-e^{\beta (\varepsilon_0-x)}}+\ldots, \label{deg-11}
\eeq
and expand the Bose factor for $x\sim \varepsilon_0$ and rearrange  to obtain the leading solution
\beq
x-\varepsilon_0= \pm\sqrt{\frac{U}{\beta}}+..\label{leading-1}
\eeq
To get the next correction we note that the right-hand side of \disp{deg-11} is expressible as $$\frac{U}{\beta (x- \varepsilon_0)(1- \half \beta (x- \varepsilon_0)+\ldots)},$$
so in the spirit of a perturbative expansion we substitute the leading result \disp{leading-1} in the denominator and expanding we get the next correction  as
\beq
x_\pm=\varepsilon_0\pm \sqrt{\frac{U}{\beta}}+\ubytwo+ {\cal O}(U^{3/2}), \label{deg12}
\eeq
giving  the first few terms in  \disp{EYLdirect-deg}.

 This expression shows that the doubly degenerate root of the U=0 system evolves into a pair of roots at non-zero U. The pair of roots
  are real for $U>0$ and complex-conjugates for $U<0$. The reader will notice   a similarity between
this result, and the results from the heuristic  argument presented in \disp{heur-1,pert-1} in the Introduction. 
\subsubsection{Satisfying the two sumrules \disp{sumrule-1,sumrule-2}}

We note that the perturbative solutions \disp{EYLdirect,EYLdirect-deg,deg12} satisfy the first sumrule \disp{sumrule-1}, one checks that terms of orders other than ${\cal O}(U)$ cancel out. The solutions \disp{EYLdirect,EYLdirect-deg,deg12} also satisfy the second sumrule \disp{sumrule-2}, here terms of  ${\cal O}(U^n)$ with $n\neq 0$    cancel out. These sumrules are therefore  of interest in numerical schemes, providing nontrivial constraints.

\section{Virtual energies from self-energy to ${\cal O}(U)$. \label{Section6}}

The self-energy of the Hubbard model on $N_s$ sites to ${\cal O}(U)$ is readily  obtained from standard perturbation theory \cite{AGD} as
\beq
\Sigma_\si(\k,i \omega_n)=\frac{U}{N_s} \sum_{\vec{p} } \frac{1}{e^{\beta(\varepsilon_{p \sib}-\mu)}+1},
\eeq
where $\sib=-\si$. Substituting into \disp{solver} we get the equation determining the virtual energy as 
\beq
\xi_{\k \si}&=& \varepsilon_{\k \si}-\ubytwo+\frac{U}{N_s}\sum_p \frac{1}{1-e^{\beta(\varepsilon_{p \sib}-\xi_{k \si}-\ubytwo)}} +{\cal O}(U^2).  
\eeq
Below we will present the numerical solutions of a simplified version
\beq
\xi_{\k \si}&=&\varepsilon_{\k \si}+\frac{U}{2 N_s}\sum_p \coth \frac{\beta}{2}(\xi_{k \si}-\varepsilon_{p \sib} ), \label{eq64}
\eeq
where we dropped a term of the ${\cal O}(U^2)$ arising from expanding the $\coth$. The dropped {   term is  added back to other second order terms in the self-energy, and the results of such a calculation to ${\cal O}(U^2)$ will be reported later \cite{SedikShastry}}. From the structure of \disp{eq64} and since $\sum_p \varepsilon_{p\sib}=0$, we see that the sum-rule \disp{sumrule-1} is automatically satisfied. 

 Let us now consider only the degenerate case where the energies $\varepsilon_{\k \si}$ shed any dependence of $\si$. This is a somewhat subtle case relative to the non-degenerate case, where continuity in $U$ can be used. For the degenerate case our treatment shows how a degenerate level can lead to a pair of distinct virtual energies at this order.
We note that the second term in \disp{eq64}, regarded as a function of $\xi$, has a pole at each $\varepsilon_p$ on the real line. Very close to $\varepsilon_a$ the equation can be approximated by retaining only the dominant term, giving two solutions $\xi^*=\varepsilon_a\pm \sqrt{\frac{U}{N_s}}$ straddling $\varepsilon_a$. This pair of  solutions can then by refined by adding in the neglected terms.
 For any non-zero $U$  we then expect $\sim$$2$$ L$ solutions. For a fixed $\k$, we want the energies to coincide with the non-interacting value $\varepsilon_{\k}$ as $U\to0$. Therefore we  pick two solutions closest to this value. {  We can use a convention where the solution with lower (higher) real part of $\xi$ is regarded as arising from spin up (down), assuming continuity when a magnetic field in the $+z$ direction is tuned to zero. }This and keeping in mind the condition of complex conjugation, i.e. $\xi$ and $\xi^*$ occuring in pairs (discussed below \disp{conventionE}),  this gives us a total of $2 L$ virtual energies, each evolving from a specific $\varepsilon_{\k}$. Solving these equations for finite systems is fairly straightforward. We display the results of solving \disp{eq64}, and compare with the exact energies found by numerical means for a small Hubbard. The partition function for a 6-site Hubbard model with periodic boundary conditions-i.e. the Benzene ring- was found using the {\em Dirac-Q} symbolic computation program\cite{DiracQ}. A comparison is presented for both repulsive and attractive signs of $U$ at $\beta=1$ (all energies are relative to the hopping $t$).

\begin{figure}[h]
\centering
\includegraphics[width=.49\columnwidth]{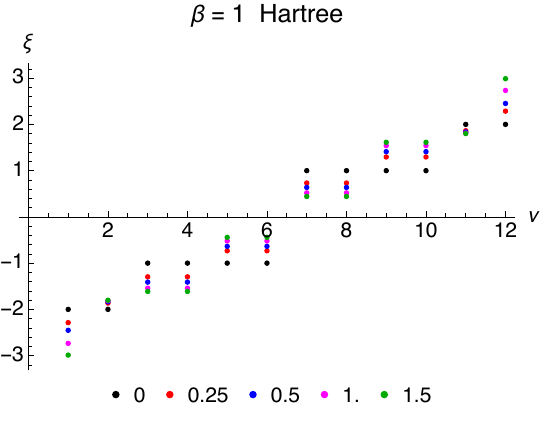}
\includegraphics[width=.49\columnwidth]{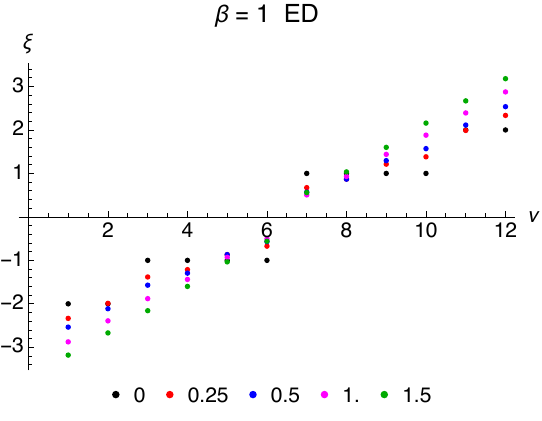}
 \caption{\footnotesize {  The virtual energies $\xi$  for the 6 site Hubbard model from the Hartree equation \disp{eq64} on  {\bf left}, and directly from the partition found from exact diagonalization (ED) on  {\bf right}, at various repulsive values of $U$ marked below the plots. The virtual energies are all real in this case.
On {\bf  left} the x-axis index ``$\nu$" combines  six increasing values of $k=\frac{2 \pi}{6}\{1,2,\ldots6\}$  with two values of the spin $\sigma=\{\uparrow,\downarrow\}$, into 12 indices $\{ (k, \uparrow),(k, \downarrow)\}$ in ascending order, using the convention discussed below \disp{eq64}. As an example $\nu=3$ corresponds to $k= \frac{2 \pi}{3}, \sigma=\uparrow$, while  $\nu=4$ corresponds to $k=\frac{2 \pi}{3}, \sigma=\downarrow$. On  {\bf right} the x-axis  label ``$\nu$"  is the sorting index of the increasing energies found from ED, which are blind to the wave vectors. We may thus overlap these indices, as in \figdisp{Figure3}}.
 The exact results show a smooth dispersion of the virtual energies-together with the expected $\xi$$\to$$-\xi$ symmetry. The approximate solutions have a similar scale and also satisfy the $\xi$$\to$$-\xi$ symmetry. We see that some of the parity related degeneracies of the non-interacting spectrum  persist in the approximate results, and are likely to be lifted in a second order calculation.
   \label{Figure2}}
 \end{figure}

In \figdisp{Figure2}  we compare the virtual energies for a few typical values of  positive $U$ found by solving \disp{eq64} in the left panel,  and by numerically finding the roots of the partition function in the right panel.
 
 In \figdisp{Figure3} we compare the exact virtual energies and the approximate ones from \disp{eq64} at four typical values of $U$. For smaller $U\leq0.25$  the results are quite close. They suggest that it is reasonable to label the numerical solutions with the $k$ values  read off from the approximate solutions.  For the larger values of $U$ the agreement is poorer, although the extremities show a closer convergence.
\begin{figure}[h]
\centering
\includegraphics[width=.49\columnwidth]{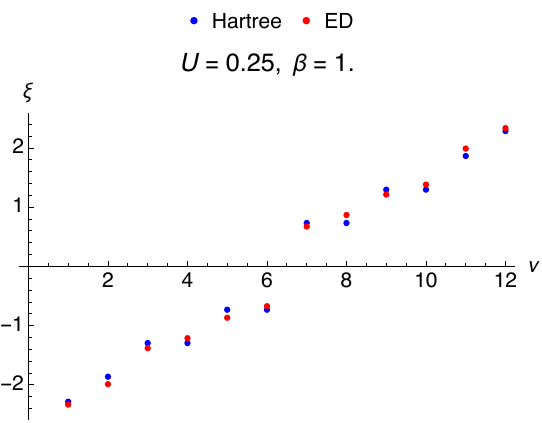}
\includegraphics[width=.49\columnwidth]{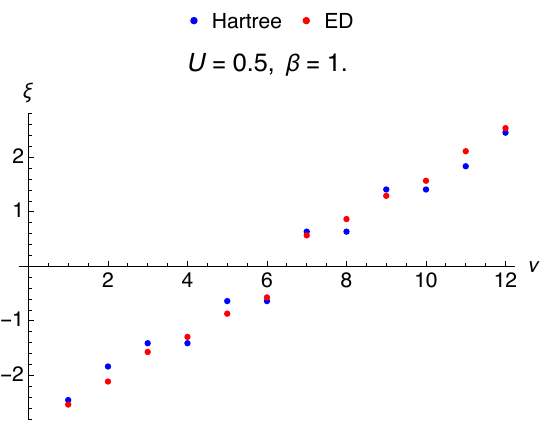}\\
\includegraphics[width=.49\columnwidth]{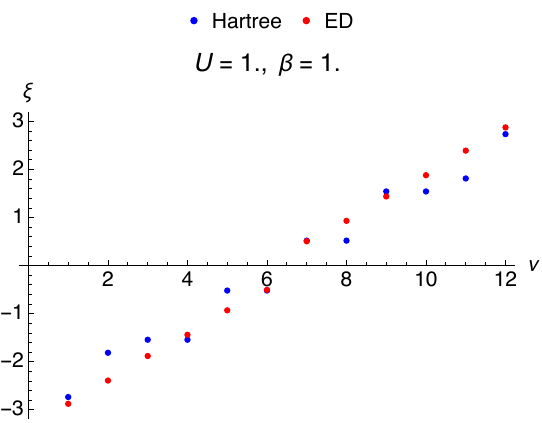}
\includegraphics[width=.49\columnwidth]{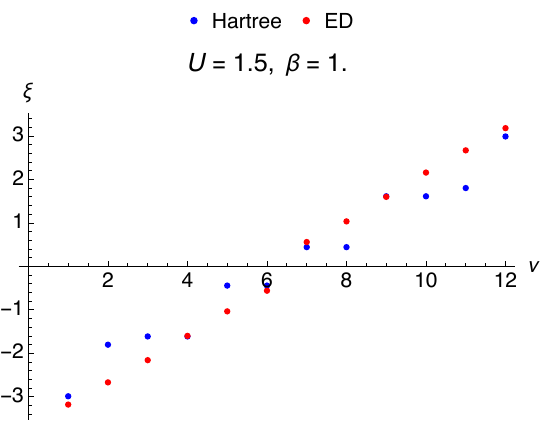}
 \caption{\footnotesize \label{Figure3} Comparison between the exact virtual energies found from the partition function zeros (red dots) and the  solutions of  the first order equations \disp{eq64} (blue dots) at four different repulsive values of $U$ marked in the plots. {  The $x$ axis label  index ``$\nu$" is as defined in \figdisp{Figure2} for the Hartree solution and also the sorting index for the energies found from  ED.}  For the displayed $U$ and $\beta$, the solutions of \disp{eq64}  can be used to compute  $U_{eff}$ (see \disp{Ueff,Ueff-2}). The pairs $(U,U_{eff})$  are $(0.25,0.149),(0.5,0.289),(1.0,0.55),(1.5,0.804)$, indicating that the first order approximation  worsens with increasing $U$. }\label{Figure3}
  \end{figure}

\begin{figure}[h]
\centering
\includegraphics[width=.46\columnwidth]{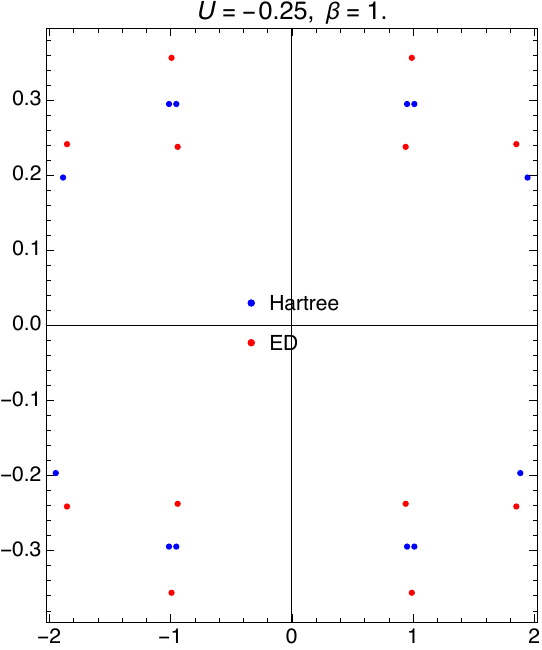}
\includegraphics[width=.46\columnwidth]{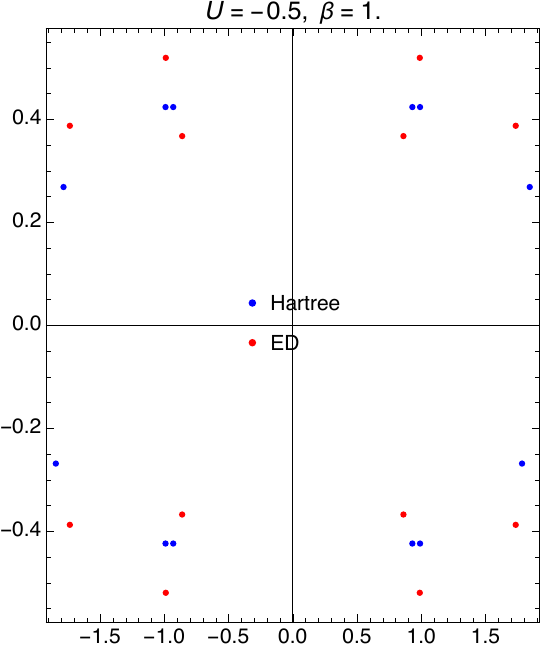}\\
\includegraphics[width=.46\columnwidth]{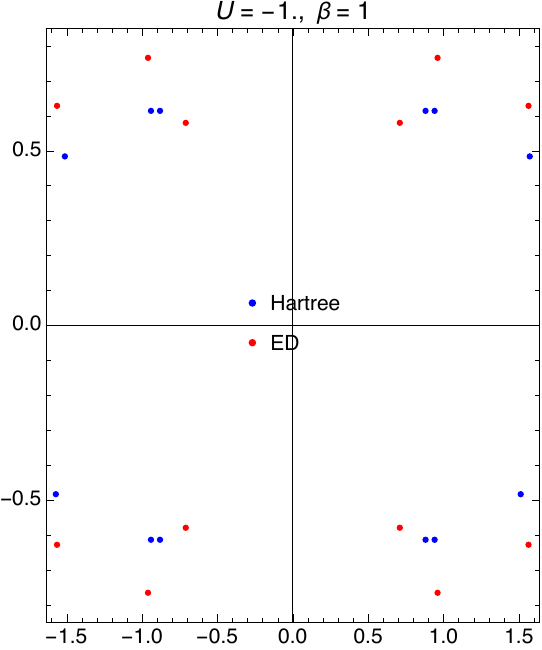}
\includegraphics[width=.46\columnwidth]{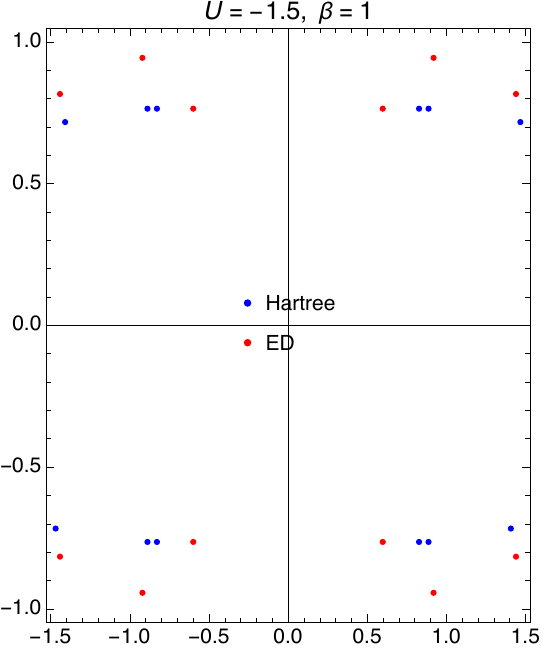}
 \caption{\footnotesize \label{Figure4} {\bf Attractive interactions} All the attractive cases lead to complex virtual energies, as expected from arguments given in the text.  Complex plots of the virtual energies with the real (imaginary) part on the x (y) axis. We
 compare  the exact virtual energies found from the partition function zeros (red dots) and the solutions of \disp{eq64} (blue dots) at four different attractive values of $U$ marked in the plots.  We picture the degenerate roots by introducing a slight displacement in the real part.
  For the displayed $U$ and $\beta$, the solutions of \disp{eq64}  can be used to compute  $U_{eff}$ (see \disp{Ueff,Ueff-2}). The pairs $(U,U_{eff})$  are $(-0.25,-.162),(-0.5,-0.344),(-1.0,-.709+0.237 i),(-1.5, -0.916+0.334 i)$, indicating  as in the repulsive case  \figdisp{Figure3}, that the first order approximation  worsens with increasing $U$.
{    The plots at larger $|U|$ show a tendency for the real parts of energies to contract. This is understandable since an attractive $U$ encourages spin-paired electrons to hop together- while discouraging spin-unpaired electron hops- so as to avoid losing their pairing energy\cite{Thierry}. The correlated hopping  bandwidth tends to shrink with increasing $|U|$.  }
 }
 \end{figure}

 In \figdisp{Figure4} we display results for the attractive case, at a few  negative values of $U$.  All virtual energies are non-real in these calculations, requiring only a negative sign of $U$. The solutions of \disp{eq64}  capture the shrinking range of the real part of the virtual energies at the extremities in all cases. The quantitative agreement with the imaginary parts is not as good. 
 
  The formation of bound (Cooper) pairs in this case is similar to that of electron-electron binding  posited in Dyson's well known argument\cite{Dyson} about the instability of quantum electrodyamics with a flipped sign of $e^2$. The resulting non-convergence of perturbation theory in field theory, is based on the instability of the ground state due to pair formation, under the {  standard} assumption of an infinite system. For the conventionally measured variables (energy, susceptibilities etc) related  effects can only be seen in the limit of a large number of particles- i.e. in the thermodynamic limit. In the present case, we see that the distinct behaviour of the virtual states between the repulsive and attractive cases, namely the energies going from real values for repulsion  to complex pairs for attraction. This striking  change is visible even for small systems.

\section{Concluding remarks \label{Section7}}
We have argued  that the  roots of the grand partition function of many-body models, such as the Hubbard model, can be found perturbatively in the interaction. Our method
gives virtual energies $\xi_{\k \si}$, which 
 are assigned (crystal) momentum and spin labels. 
 The condition  \disp{solver}, can be used to systematically estimate the virtual energies $\xi_{\k \si}$, starting from the non-interacting values $\xi_{\k \si}\vert_{U\to 0}\to \varepsilon_\si(\k) $. We have illustrated this idea with two examples (i) { The single site Hubbard model}, by solving for the  thermodynamically derived virtual energies  by perturbation theory to second order in U and compared with expansion in $U$ of the exact values that are readily found (ii) { the Benzene molecule, i.e. a 6-site Hubbard ring} by solving for the virtual energies in a 6-site Hubbard model by  first  order perturbation theory in $U$  of either sign and compared with the exact numerical results.

For repulsive U, the virtual energies are  real in the lowest order  calculations. Assuming that a substantial fraction of the virtual roots are real at low $T$ in certain energy ranges, their level density should control the compressibility, heat capacity  and susceptibility for the corresponding densities. In this sense
  the methods developed here may be expected 
to connect with the Landau Fermi liquid and its many variants\cite{Leggett}. Towards this end it seems important to study the virtual energy distribution for larger systems, and to study its variation with the strength of the interaction $U$.
For the case of attractive interaction, the results presented here give  non-real virtual energies, already to first order in U and for very  small lattice sizes.   Further applications of these ideas  will be published in a forthcoming publication\cite{SedikShastry}. 
 
 \section{Acknowledgements} I thank Professor Elliott   Lieb for helpful discussions   in 2018  at Princeton. I also thank    Muhammad Sedik for  helpful comments.

\appendices
 \section{Zeros from the Luttinger-Ward formula \label{Section8}}

In this section we discuss another expression for the partition function of a many-body system found in  the well-known work of Luttinger and Ward (LW)\cite{Luttinger-Ward}.
 LW found a formally exact expression for the thermodynamic potential $\Omega$ (and therefore the partition function) {  in terms of the Greens function $G_\sigma(\k,i \omega_r)$ },  from perturbation theory to all orders in the coupling. The LW result follows from a  clever rearrangement of the thermal perturbation theory, whose convergence is implicitly assumed to be sufficiently ``benign''. Their  result for the grand potential $\Omega$ 
\beq  \Omega= - \invbeta \log Z(\mu)  \eeq
can be written in terms of the partition function in the following way:
  \beq
  Z(\mu)= Z_A(\mu)\times Z_B(\mu)\times Z_C(\mu) \label{part-LW}
 \eeq
where
\beq
 \log Z_A(\mu)& =&  \sum_{\k \, r \si}  e^{i \omega_r 0^+} \; \log [- G_\si^{-1}(\k, i \omega_r) ]   \label{logZA}
 \eeq
\beq
 \log Z_B(\mu)& =&  \sum_{\k \, r \si}  e^{i \omega_r 0^+} \{ G_0^{-1}(\k,i \omega_r) G_\si(\k,i \omega_r) -1\}   \label{logZB}
 \eeq
and
 \beq
 \log Z_C(\mu)=-  \sum_{\nu\geq 1}   \sum_{\k \,  r \si} e^{i \omega_r 0^+}  \frac{1}{2 \nu} \widetilde{\Sigma}^{(\nu)}_\si(\k,i \omega_r)  \; G_\si(\k, i \omega_r)\label{logZC}
 \eeq
 where $G_0$ is the non-interacting Greens function, $\widetilde{\Sigma}^{(\nu)}_\si(\k,i \omega_r)$  is the total $\nu$th order skeleton self-energy part, where the order of the diagram $\nu$ counts all explicit powers of the interaction, {  in this case} U.
 In these expressions  the term  $e^{i \omega_r 0^+}$ is required for convergence of the sum. Note that $Z_B$ and $Z_C$ vanish in the limit of free Fermions. Readers might recognize that $\log Z_C$ \disp{logZC} is the   Luttinger-Ward functional of $G$, with the property that its 
 functional derivative gives the exact self-energy, and thereby $\Omega$ is a stationary functional of $G_{\k \si}$.

 From \disp{logZA} we obtain the product form
 \beq
 Z_A(\mu)&=& \prod_{\k  \si \, r}  e^{i \omega_r 0^+} \;  \{- G_\si^{-1}(\k, i \omega_r) \} \label{ZA-prod}, \nn \\
 &=& \prod_{\k \si} (-1) \Psi_{\k \si}(\mu)  \times \left( \prod_{\k  \si \, r\neq 0 }  e^{i \omega_r 0^+} \;  \{- G_\si^{-1}(\k, i \omega_r|\mu) \} \right)   \eeq
 using \disp{Psi}.  Indeed the complete partition function $Z_0$ of the free Fermi theory is of this form with $G_0$ replacing $G$.  It is seen from this expression that  $Z_A$ (and hence $Z(\mu)$) contains the factor
 $\Xi(\mu)$ given in \disp{our-main}, and their quotient  is expressible as  a functional of $G$. 
The  quotient is required to be a holomorphic function of $\mu$ 
 in the strip ${\cal S}$, with no other zeros.
This result  not easy to prove directly, but must be true if all the zeros are given by the function $\Xi(\mu)$ given in \disp{our-main}.

\end{document}